\documentstyle[12pt,a4]{article}
\begin{document}
\setlength{\textheight}{9.4in}
\setlength{\topmargin}{-0.6in}
\setlength{\oddsidemargin}{0.2in}
\setlength{\evensidemargin}{0.5in}
\renewcommand{\thefootnote}{\fnsymbol{footnote}}
\begin{center}
{\Large {\bf The  Wess--Zumino term and quantum tunneling}}
\vspace{0.8cm}

{J.--Q. Liang\raisebox{0.8ex}{\small a,b}\footnote[1]
{DAAD-K.C.Wong--Fellow, Email:liangjq@physik.uni-kl.de, liangjq@sun.ihep.ac.cn},
L. Maharana\raisebox{0.8ex}{\small a,c}\footnote[2]{DAAD--Fellow,
Email:maharana@physik.uni-kl.de, lmaharan@iopb.stpbh.soft.net}
and H.J.W. M\"uller--Kirsten\raisebox{0.8ex}{\small a}\footnote[3]{Email:mueller1@physik.uni-kl.de}\\

\raisebox{0.8 ex}{\small a)}{\it Department of Physics, University of Kaiserslautern, D--67653
Kaiserslautern, Germany}

\raisebox{0.8 ex}{\small b)}{\it Institute of Theoretical Physics, Shanxi University,
Taiyuan, Shanxi 030 006, P.R. China}

\raisebox{0.8 ex}{\small c)}{\it Department of Physics, Utkal University, Bhubaneswar 751 004,
India}}
\end{center}
\vspace{0.3cm}

{\centerline {\bf Abstract}}
\noindent
The significance of the Wess--Zumino term in spin tunneling is explored,
and
a formula is established for the splitting of energy levels
of a particle with large fermionic spin as an applied
magnetic field is switched on. 
\begin{center}
\line(1,0){50}
\end{center}
\vspace{0.5cm}

The socalled Wess--Zumino term in an action integral has
increasingly been realised to be a quantity
of considerable significance and with far--reaching physical consequences.  The original
work usually cited, ref.\cite{1}, is not very illuminating in this respect.  The truly deep
content of the quantity has, however, been uncovered particularly in ref.\cite{2} in
connection with the quantisation of the Skyrme model (and
the identification of the integer contained in the 
coefficient of the Wess--Zumino term with the number of colours),
and in ref.\cite{3}, for instance, in the very different
area of macroscopic quantum tunneling.
Here we consider the latter case.  We show why the term -- which arises from
the quasiclassical consideration which here implies the
use of the coherent state representation -- is a Wess--Zumino
term, we explore some of its characteristics and establish a
formula for the level splitting of a particle with large spin $S$
(integral or half--integral) in an applied magnetic field.  The fermionic suppression of the
level splitting in special cases has been established earlier (cf. refs.\cite{3,4,5}). Our
aim here is to extend to the half--integral case 
the formula obtained earlier for integral $S$
by path integral methods \cite{6} and from Schr\"odinger theory
\cite{7,8} and in the presence of an applied magnetic field.
In many one--dimensional cases Schr\"odinger theory is not only useful
for comparison purposes, but is in particular simpler than
the path integral method for transitions at the level
of excited states, as a comparison shows (cf. e.g. \cite{8} and \cite{9,10,11}).

The consideration of spin systems is basically a discrete problem
but in a quasiclassical treatment its mapping into a
continuous system by first replacing spin operators $\hat{S}$ by classical
vectors of length $S$ has turned out to be an
exceedingly useful procedure for the construction
of models.  Thus the spin Hamiltonian with
two different anisotropy axes and
without application of an external magnetic field,
\begin{equation}
\hat{{\cal H}} = - k_x{\hat S}^2_x + k_z{\hat S}^2_z
\label{1}
\end{equation}
($k_x, k_z>0$) is related to the classical Hamiltonian ($0\leq\theta\leq\pi, 0\leq\phi\leq2\pi)$
\begin{equation}
{\cal H} = -k_xS^2\sin^2\theta\cos^2\phi + k_zS^2\cos^2\theta
\label{2}
\end{equation}
with two degenerate minima at $\theta=\frac{\pi}{2}, \phi = 0,\pi$.
Quantum mechanically the spin can tunnel between these minima.  The quasiclassical
formulation implies the use of coherent (i.e. minimal uncertainty) states
in the evaluation of the path integral.  We impose the condition that
these states be single--valued under the transformation $\phi\rightarrow \phi+2\pi$
in order to force phase effects into the action (thus leading to the Wess--Zumino
term).  The coherent states can be shown to be only
asymptotically (i.e. for $S\rightarrow\infty$)
orthogonal, and it is their overlap at neighbouring Euclidean time steps which
gives rise to a phase, the Wess--Zumino term in the Euclidean action $S_E$\cite{3}.

If we consider a single spin, the action is an integral over Euclidean time $\tau$
(and one obtains a level splitting).
If we consider a large number of individual spins which align
at low temperature, we can view the system as a single
large spin, and the action is in addition to
the integral over $\tau$ a sum over lattice sites (with
separation $a$) or an integral over
a spatial coordinate $x$ (resulting in energy bands due to
spatial translational invariance).  In this case the Hamiltonian
has in addition a Heisenberg term, i.e.
\begin{equation}
\hat{{\cal H}} =\frac{1}{aS^2}\left[ - J\sum_{i,\rho}{\hat S}_i\cdot{\hat S}_{i+\rho}-{K}_
x\sum_i({\hat S}^x_i)^2 + 
{K}_z\sum_i({\hat S}^z_i)^2\right]
\label{3}
\end{equation}
In either case the equivalent effective quasiclassical continuum action $S_E$ in Euclidean time
obtained with the
use of coherent states assumes the general form \cite{3}
\begin{equation}
S_E=S_{WZ} + \int^{\beta}_0d\tau L,\;\;\; S_{WZ} = i\gamma\int^{\beta}_0d\tau\dot{\psi},\;\;
 L=\frac{1}{2}M(\psi)\dot{\psi}^2 + V(\psi)
\label{4}
\end{equation}
where $\dot{\psi} = d\psi/d\tau$, and the constant $\gamma$, the effective mass $M(\psi)$
and the periodic potential $V(\psi)$ are quantities with model dependent parameters.
In the single spin case referred to above 
$\psi$ is simply the azimuthal angle $\phi$ with variation from one potential
minimum at $0$ or $\pi$ to the other at
$\pi$ or $0$ respectively as $\tau$ varies from $0$ to $\beta$, and $\gamma = S$
(the polar angle $\theta$ can conveniently be
chosen to be $\frac{\pi}{2}$). 
The Wess--Zumino action $S_{WZ}$ does not only depend on the
boundary values $\phi (0), \phi(\beta)$ but also on
the path.  The endpoint $\phi(\beta)$ can be reached
on the circle of unit radius along the right hand semicircular arc $a_r$ in
the positive direction of ${\bf e}_{\phi}$ or along
the left hand arc $a_l$ in the opposite direction.
This direction--dependence implies a handedness or chirality
$C_{\pm}$ in $S_{WZ}$, i.e.
\begin{equation}
S_{WZ}=iS\pi C_{\pm}, \;\;\; C_{\pm}=\frac{1}{\pi}\left[\int_{a_{r,l}}
d\tau\dot{\phi}\right]^{\beta}_0 = \frac{1}{\pi}\left[a_{r,l}\right]^{\beta}_0
\label{5}
\end{equation}
which in the present case (without applied magnetic field) implies
$C_{\pm} = \pm1$.
The effective mass $M(\phi)$ and potential $V(\phi)$ are found
to be (with the conventions of eq.({\ref{1})) \cite{6,7,8}
\begin{equation}
M(\phi) = \frac{1}{2k_z\left(1+\frac{k_x}{k_z}\cos^2\phi\right)},\;\;\; V(\phi)
=- k_xS^2\cos^2\phi
\label{6}
\end{equation} 
The Euclidean time equation of motion can be shown to possess periodic instanton
solutions \cite{5,10} which reduce in the limit $\frac{k_x}{k_z}\rightarrow 0$ to the
periodic and vacuum instanton solutions
of sine--Gordon theory. It is only the vacuum configuration which saturates the
Bogomol'nyi inequality which can be constructed from the expression of its (positive) energy.  
Since in most cases in the context of spin tunneling the relevant pseudoparticle
configurations are not the vacuum ones, this inequality is not of primary importance here.
In the second, multiple spin, case referred to above the periodic instanton configurations appear
as static (i.e. $\tau$--independent) solutions of the 1--dimensional equation of motion.  The
translational invariance of the equation then implies the existence of a dynamical collective
coordinate $\chi(\tau)$. Reexpressing the effective
action entirely in terms of this collective coordinate, one again arrives at an expression
like that of eq.(\ref{4}) with $\psi$ replaced by $\chi$ \cite{3}.  
In this case $S_{WZ}=iS\pi C_{\pm}$ where now $S=Ns|\chi(\beta)-\chi(0)|/a$, $N$ being the number
of spins and $s$ the spinvalue of one of them.
It is the action integral $S_E$ of eq.(\ref{4}) which defines the
basic  model theory of the ferromagnetic sample we 
consider.

That $S_{WZ}$ is the action of a Wess--Zumino term as in
field theory follows from its properties and the
fact that we can rewrite it in the defining form of ref. \cite{2}.  We first observe that as a total
derivative it does not contribute to
the equation of motion. 
Clearly we can rewrite $C_{\pm}$ as
\begin{equation} 
C_{\pm}=-\frac{i}{\pi}\int_{a_{r,l}} d\tau [g^{\dagger}\frac{\partial}{\partial\tau }g],
\;\;\; g = e^{i\psi (\tau)} 
\label{7}
\end{equation}
thus demonstrating its
appearance as a Wess--Zumino term
(the higher dimensional form contains more factors like the $\tau$--dependent one here).
We observe that without the Wess--Zumino term
the Lagrangian density is separately invariant under the
global replacements $\tau\rightarrow -\tau$
and $\psi\rightarrow -\psi$, but with it only under the combination of both.  Thus the
Wess--Zumino term restricts the symmetry of the Lagrangian. As illustrated with
the help of a simple model in ref. \cite{2} it is precisely
such a condition
which gives rise to a Wess--Zumino term. The chirality associated with
this term leads
to a
gauge potential, ${\bf A}$, which in the present case is simply
a unit vector ${\bf e}_{\phi}$ around the unit circle $S^1=a_l\cup  a_r$
in the $xy$--plane, travelling either
in the clockwise sense or in the anticlockwise sense through an angle  $2\pi$.
Thus
\begin{equation}
\int_ad\tau\dot{\phi}(\tau)=\int_a{\bf e}_{\phi}\cdot \mbox{\boldmath$d\phi$}
\equiv\int_a{\bf A}\cdot \mbox{\boldmath$d\phi$}
=\int_a \mbox{\boldmath$A$}\cdot\dot{\mbox{\boldmath$\phi $}}\, d\tau
\label{8}
\end{equation}
and
$$
e^{S_{WZ}(a_r\cup a_l)}=1,\;\;S_{WZ}(a_r\cup a_l)=2iN\pi
$$
($N$ an integer), which is the condition corresponding to the Dirac charge quantisation
condition (and, in fact, in ref. \cite{2} the number of colours appears in the coefficient
of the Wess--Zumino term in a similar way).
Thus also
$$
e^{S_{WZ}(a_r)}=e^{-S_{WZ}(a_l)}
$$
One should note, that even
in the 2--dimensional case of the multi--spin case above, the Wess-Zumino term is
the same one-dimensional one as above.  A 2-dimensional generalisation of the above,
say to $\tau$ and $x$, would not only be zero (in view of the
necessary $\epsilon_{\tau x}$ in front, but in addition would not satisfy such a
symmetry requirement). 
Another way to see that the Wess--Zumino term does not affect the
{\it mechanical} energy is to write down the Hamiltonian.  With the definition
$p_{\phi}=\frac{\partial(L+iS\dot{\phi})}{\partial{\dot{\phi}}} = M(\phi)\dot{\phi}+iS$
of the conjugate momentum (the first part being the mechanical
momentum), one finds immediately that in the Hamiltonian
$H = p_{\phi}\dot\phi - (L+iS\dot{\phi})$ the Wess--Zumino contribution
drops out and one obtains
$$
H = \frac{1}{2}M(\phi) {\dot{\phi}}^2 - V(\phi)
$$
as expected. Expressed in terms of canonical variables the Hamiltonian, of course, assumes
the form as in gauge theory, i.e.
$$
H = \frac{(p_{\phi}-iS)^2}{2M(\phi)} - V(\phi)
$$ 
This also shows that the Wess--Zumino term does not affect the classical
energy of a pseudoparticle solution of the equations of motion.

The problem becomes more complicated if we also allow
an externally applied magnetic field ${\cal B}$, e.g. perpendicular to the
easy axis $x$, which can be taken into account by adding to the
Hamiltonian of eq.(\ref{1}) the contribution $-h\hat{S}_y$, where
$h = g\mu_B{\cal B}$.  Then
\begin{equation}
\hat{\cal H}= -k_x\hat{S}^2_x + k_z\hat{S}_z^2 - h\hat{S}_y
\label{9}
\end{equation}
In the corresponding quasiclassical continuum representation the new
contribution implies that $M$ and $V$ (with $\psi\rightarrow\phi$)
change to (cf. e.g. ref.\cite{6})
\begin{eqnarray}
M(\phi)&=&\left[2k_z+2k_x\cos^2\phi +\frac{h\sin\phi}{S+\frac{1}{2}}\right]^{-1},\nonumber\\
V(\phi)&=&-\left[k_xS(S+1)\cos^2\phi + h(S+\frac{1}{2})\sin\phi\right]
\label{10}
\end{eqnarray}
For integral values of $S$ this case has been considered earlier in refs.\cite{6,8,12}.  Our                      
interest here concerns the general case of $S$ integral or half--integral and
the establishment of a formula for the level splitting particularly in the latter case.  We therefore
follow ref.\cite{8}. Instead of using the path integral method
as in refs. \cite{6,12}, we consider the
equivalent Schr\"odinger theory and proceed as in refs.\cite {7,8}.  The procedure employed there
exploits the known level splitting for a periodic potential with identical barriers
on either side of one of the wells of the potential.
The appropriate  one--dimensional Schr\"odinger equation with mass $\frac{1}{2}$ and
potential $V(\phi) = 2h_m^2\cos2\phi$ with $h_m^2$ assumed to be
large (high barriers) leads to a splitting of the $n$--th asymptotically--single
oscillator level given by the eigenvalue difference (cf. \cite{7,8} and references therein)
\begin{equation}
\triangle_{q_0=2n+1}= \frac{2(16h_m)^{q_0/2+1}e^{-4h_m}}
{(8\pi)^{1/2}[\frac{1}{2}(q_0-1)]!}\left(1+O(\frac{1}{h_m})\right)
\label{11}
\end{equation}
This result has also been obtained by path integral methods \cite{11}.
Identifying parameters with the present case, one finds (cf.\cite{8})
\begin{equation}
h_m^2=\frac{k_x}{4k_z}S(S+1)(1-a^2), \;\; a=\frac{h}{2k_x(S+\frac{1}{2})},\;\; b=\frac{k_z}{k_x}
\label{12}
\end{equation}
The effect of the applied
magnetic field and so $h$ is to shift the minima of the periodic potential slightly away
from an integral value of $\pi$, and also to
make successive barriers of the potential asymmetric, i.e. the barrier on one side of
a minimum differs in size from that on the other side.
Choosing the potential minimum at $\phi = \pi -\arcsin a$ as our reference minimum
as in refs. \cite{8,12},
the adjacent minima to the left and to the right are located angles
$$
\phi_l=\arcsin a, \;\;\phi_r=  2\pi +\arcsin a
$$
implying arc lengths
$$
a_l=-\pi+2\arcsin a, \;\;\; a_r = \pi + 2\arcsin a
$$
respectively, so that the distance from the minimum to the
left of the reference minimum to that on
the right is $-2\pi$.  
Correspondingly the
periodic instantons in these barriers also differ
(the explicit expressions are given in refs. \cite{8,12}).  The actions
${\cal A}_{l,r}$ of these instantons enter
the argument of the exponential contained in the level splitting whereas the
factor in front is determined by the matching of different branches of the
wave function in domains of overlap. Our present consideration differs from that
in ref. \cite{8} in that we now also wish to take into account the Wess--Zumino
term.  Thus we have to add to the action of either periodic instanton a contribution
$iS\pi C_{\pm}$. This means that the tunneling exponential, which in the fieldless
case ($a=0$) is 
$$
e^{-\frac{2S}{\sqrt{b}}}
$$ 
must be replaced by half the sum of two exponentials with arguments
$-{\cal A}_r-iS\pi C_+$ and $-{\cal A}_l-iS\pi C_-$ respectively, i.e.
\begin{eqnarray}
\frac{1}{2}e^{-\frac{2S}{\sqrt{b}}\left(\sqrt{1-a^2} + a\arcsin a\right)}
&\bigg[&e^{-i(\pi +2\arcsin a) S}.e^{a(s+\frac{1}{2})\frac{\pi}{\sqrt{b}}}\nonumber\\
&+&e^{i(\pi-2\arcsin a) S}.e^{-a(s+\frac{1}{2})\frac{\pi}{\sqrt{b}}}\bigg]\nonumber\\   
=e^{-\frac{2S}{\sqrt{b}}\left(\sqrt{1-a^2}+a\arcsin a\right)}&.&
e^{-i2S\arcsin a}.\cos 
\left(S\pi-\frac{ia(S+\frac{1}{2})\pi}{\sqrt{b}}\right)
\label{13}
\end{eqnarray}
The $a$--dependent parts of the phase appearing
here
are sometimes
called 
``Aharonov--Bohm'' contributions \cite{13}.
With the appropriate replacement in eq. (\ref{12}) the level splitting in this
general case becomes
\begin{eqnarray}
\triangle_{q_0=2n+1} &=& \frac{2(16h_m)^{q_0/2+1}}{(8\pi)^{1/2}[\frac{1}{2}(q_0-1)]!}.
e^{-\frac{2S}{\sqrt{b}}(\sqrt{1-a^2}+a\arcsin a)}\nonumber\\
&.&\left|e^{-i2S\arcsin a}\cos\left(S\pi-\frac{ia(S+\frac{1}{2})\pi}{\sqrt{b}}\right)\right|
\label{14}
\end{eqnarray}
We take the modulus of the phase factor not only because the energy must be
positive, but also in view of the way the factor has to be handled in the
path integral method (cf. ref. \cite{3}). For integral values of $S$ this formula reduces to
\begin{equation}
\triangle_{q_0=2n+1} = \frac{2(16h_m)^{q_0/2+1}}{(8\pi)^{1/2}[\frac{1}{2}(q_0-1)]!}.
e^{-\frac{2S}{\sqrt{b}}(\sqrt{1-a^2}+a\arcsin a)}\cosh\left(\frac{a(S+\frac{1}{2})\pi}{\sqrt{b}}\right)
\label{15}
\end{equation}
This is the formula obtained in \cite{8} in agreement with the results in \cite{6}.  For
half--integral values of $S$ and applied field zero, i.e. $a = 0$, the splitting
vanishes -- in agreement with the socalled Kramer's degeneracy.  The really
interesting case is that of half--integral values of $S$ and magnetic field
unequal zero. In this case we obtain  
\begin{equation}
\triangle_{q_0=2n+1} = \frac{2(16h_m)^{q_0/2+1}}{(8\pi)^{1/2}[\frac{1}{2}(q_0-1)]!}.
e^{-\frac{2S}{\sqrt{b}}(\sqrt{1-a^2}+a\arcsin a)}\sinh\left(\frac{a(S+\frac{1}{2})\pi}{\sqrt{b}}\right)
\label{16}
\end{equation}
which is a plausible result because it vanishes completely for vanishing
magnetic field. Comparing eq.(\ref{15}) with eq.(\ref{16}) we see that a very weak
but nonzero magnetic field for which the level splitting
increases linearly with the magnetic field is
indicative of a macroscopic fermionic state. It would be
interesting to see this observed.

\vspace{2cm}



\begin{thebibliography}{99}
\bibitem{1} J.Wess and B. Zumino, Phys. Lett. {\bf B37} (1971) 95.
\bibitem{2} E. Witten, Nucl. Phys. {\bf B223} (1983) 422, 433.
\bibitem{3} H.--B. Braun and D. Loss, {\it Spin parity effects and macroscopic quantum
coherence of Bloch walls}, in L.Gunther and B. Barbara (eds.), Quantum Tunneling
of Magnetisation -- QTM'94, (Kluver Academic, 1995) 319.
\bibitem{4} D. Loss, D.P. DiVincenzo and G. Grinstein, Phys. Rev. Lett. {\bf 69} (1992) 3232.
\bibitem{5} J.--Q. Liang, H.J.W. M\"uller--Kirsten and Jian--Ge Zhou, Z. Phys. {\bf B102}
(1997) 525 (note that the factors 3 in formulae (30), (32) are
misprints).
\bibitem{6} M. Enz and R. Schilling, J.Phys.{\bf C19} (1986) 1765, L711.
\bibitem{7}J.--Q. Liang, H.J.W. M\"uller--Kirsten and J.M.S. Rana, Phys. Lett.{\bf A231}
(1997) 255.
\bibitem{8} J.--Q. Liang, H.J.W. M\"uller--Kirsten, A.V. Shurgaia and F. Zimmerschied,
Phys. Lett. {\bf A237} (1998) 169.
\bibitem{9} J.--Q. Liang, H.J.W. M\"uller--Kirsten, Jian--Ge Zhou and F.C. Pu,
Phys. Lett. {\bf A228} (1997) 97.
\bibitem{10} J.--Q. Liang, Y.--B. Zhang, H.J.W. M\"uller--Kirsten, Jian--Ge Zhou, F.
Zimmerschied and F.--C. Pu, Phys. Rev. {\bf B57} (1998-I) 529.
\bibitem{11} J.--Q. Liang and H.J.W. M\"uller--Kirsten, Phys. Rev. {\bf D51} (1995) 718.
\bibitem{12} J.--Q. Liang, H.J.W. M\"uller--Kirsten, Jian--Ge Zhou
and F. Zimmerschied, Phys. Lett. {\bf B393}
(1997) 368.
\bibitem{13} E. M. Chudnovsky and D.P. DiVincenzo, Phys. Rev.{\bf B48} (1993)
10548; N. V. Prokof'ev and P.C.E. Stamp, QTM'94 (ref.\cite{3} above), p. 347.


\end{thebibliography}
\end{document}